\documentclass[]{emulateapj}

\begin{document}

\title{Exploring the convective core of the hybrid $\delta$ Scuti-$\gamma$ Doradus star CoRoT 100866999 with asteroseismology}
\author{Xinghao Chen\altaffilmark{1,2}, Yan Li \altaffilmark{1,2,3,4}, and Xiaobin Zhang\altaffilmark{5}}

\altaffiltext{1}{Yunnan Observatories, Chinese Academy of Sciences, P.O. Box 110, Kunming 650216, China; chenxinghao@ynao.ac.cn; ly@ynao.ac.cn}
\altaffiltext{2}{Key Laboratory for Structure and Evolution of Celestial Objects, Chinese Academy of Sciences, P.O. Box 110, Kunming 650216, China}
\altaffiltext{3}{University of Chinese Academy of Sciences, Beijing 100049, China}
\altaffiltext{4}{Center for Astronomical Mega-Science, Chinese Academy of Sciences, 20A Datun Road, Chaoyang District, Beijing, 100012, China}
\altaffiltext{5}{Key Laboratory of Optical Astronomy, National Astronomical Observatories, Chinese Academy of Sciences, Beijing, 100012, China; xzhang@bao.ac.cn}
\begin{abstract}
We computed a grid of theoretical models to fit the $\delta$ Scuti frequencies of CoRoT 100866999 detected earlier from the CoRoT timeserials. The pulsating primary star is determined to be a main sequence star with a rotation period of $4.1^{+0.6}_{-0.5}$ days, rotating slower than the orbital motion. The fundamental parameters of the primary star are determined to be $M$ = $1.71^{+0.13}_{-0.04}$ $M_{\odot}$, $Z=0.012^{+0.004}_{-0.000}$, $f_{\rm ov}$ = $0.02^{+0.00}_{-0.02}$, $T_{\rm eff}$ = $8024^{+249}_{-297}$ K,  $L$ = $11.898^{+2.156}_{-1.847}$ $L_{\odot}$, $\log g$ = $4.166^{+0.013}_{-0.002}$, $R$ = $1.787^{+0.040}_{-0.016}$ $R_{\odot}$, and $X_{\rm c}$ = 0.488$^{+0.011}_{-0.020}$, matching well those obtained from the eclipsing light curve analysis. Based on the model fittings, $p_1$ and $p_5$ are suggested to be two dipole modes, and $p_3$, $p_4$, $p_6$, and $p_7$ to be four quadrupole modes. In particular, $p_4$ and $p_7$ are identified as two components of one quintuplet. Based on the best-fitting model, we find that $p_1$ is a g mode and the other nonradial modes have pronounced mixed characters, which give strong constraints on the convective core. Finally, the relative size of the convective core of CoRoT 100866999 is determined to $R_{\rm conv}/R$ = $0.0931^{+0.0003}_{-0.0013}$.
\end{abstract}

\section{Introduction}
The $\delta$ Scuti (Campbell $\&$ Wright 1900) and $\gamma$ Dor variables (Kaye et al. 1999)  are two classes of late A- and early F-type stars. $\delta$ Scuti stars mainly pulsate in low-order radial and nonradial modes with typical periods in the range of 0.02-0.25 days (Breger 2000), which are driven by the $\kappa$ mechanism (Baker $\&$ Kippenhahn 1962, 1965; Zhevakin 1963; Li $\&$ Stix 1994) operating in the second partial ionization zone of helium (Chevalier 1971; Dupret et al. 2004; Grigahc{\`e}ne et al. 2005). Due to the low radial order, oscillations of $\delta$ Scuti stars are not in the asymptotic regime, while their pulsation patterns still exhibit regularities (e.g., Papar\'o et al. 2013, 2016). These regularities also include the scaling relation between the large frequency separation and the mean density of the star (Su\'arez et al. 2014; Garc\'ia Hern\'andez et al. 2015, 2017), the amplitude modulation (Bowman \& Kurtz 2014; Barcel\'o Forteza et al. 2015; Bowman et al. 2016), and the empiric relation between the frequency of the maximum oscillation power $\nu_{\rm max}$ and the effective temperature $T_{\rm eff}$ (Barcel\'o Forteza et al. 2018). $\gamma$ Dor stars pulsate in g-mode oscillations driven by the convective blocking mechanism, which operates in the outer convective zone (Guzik et al. 2000; Dupret et al. 2004, 2005; Grigahc{\`e}ne et al. 2005). Their periods are relatively longer, i.e., between 0.3 days and 3 days. The g modes of $\gamma$ Dor stars and their departure from the constant period spacing allow us to probe the interiors of the star, such as chemical mixing (Miglio et al. 2008) and rotation (Van Reeth et al. 2015, 2016, 2018).

 In the Hertzsprung-Russel diagram, the instability strips of $\delta$ Scuti and $\gamma$ Dor stars largely overlaps with each other (Balona 2011; Henry et al. 2011; Uytterhoeven et al. 2011; Xiong et al. 2016). Xiong et al. (2016) showed that most of the pulsating variables in the $\delta$ Sct-$\gamma$ Dor instability strip are very likely hybrids that pulsate in both p and g modes. Moya et al. (2017) found that the boundary of $\delta$ Scuti and $\gamma$ Dor pulsations depends on the temperature. The first such hybird star was detected from  the ground by Henry $\&$ Fekel (2005). Thanks to the space missions MOST (Walker et al. 2003), CoRoT (Baglin et al. 2006), and Kepler (Borucki et al. 2010), the hybrid behavior is found to be common for A- and F-type stars (Grigahc$\grave{\rm e}$ne et al. 2010; Balona et al. 2015), and a large number of hybrid $\delta$ Sct-$\gamma$ Dor pulsators have been detected and precisely observed, such as CoRoT 1057330033 (Chapellier et al. 2012), KIC 11145123 (Kurtz et al. 2014), and KIC 9244992 (Saio et al. 2015). The hybrid pulsators are very important and promising objects for the study of the stellar structure, since presence of p- and g-modes allow us probe properties of the star from the envelope to the core.

CoRoT 100866999 is an eclipsing binary  observed from May 16 to October 5 in 2007 ($\Delta T = 142$ days) during CoRoT's first long run targeting the Galactic center (LRc01).  Sarro et al. (2013) obtained an effective temperature of 7700 $\pm$ 400 K using low-resolution spectroscopy observed with the Giraffe multi-object spectrograph installed at VLT at ESO in Chile. Chapellier $\&$ Mathias (2013) analysed the eclipsing light curve, and determined physical parameters of the pulsating primary star as $M_1$ = 1.80 $\pm$ 0.2 $M_{\odot}$, $R_1$ = 1.90 $\pm$ 0.2 $R_{\odot}$, $\log g_1$ = 4.1 $\pm$ 0.1, $T_{\rm eff, 1}$ = 7300 $\pm$ 2500 K, and  those of the secondary star as $M_2$ = 1.1 $\pm$ 0.2 $M_{\odot}$, $R_2$ = 0.9 $\pm$ 0.2 $R_{\odot}$, $\log g_2$ = 4.6 $\pm$ 0.1, $T_{\rm eff,2}$ = 5400 $\pm$ 430 K, respectively. Besides, they extracted 8 independent $\delta$ Scuti frequencies in the domain [16.25; 26.67] days$^{-1}$ and 63 independent $\gamma$ Dor frequencies in the domain [0.30; 3.61] days$^{-1}$ (Table 2 of Chapellier $\&$ Mathias (2013) ). Moreover, they identified 22 $\gamma$ Dor frequencies as g modes of $\ell = 1$ with successive radial orders due to their nearly constant period interval.

S$\acute{\rm a}$nchez Arias et al. (2017) had carried out detailed asteroseismic modelling for CoRoT 100866999. However, their asteroseismic parameters deviate from the results of the eclipsing analysis given by Chapellier $\&$ Mathias (2013). 
In general, the component stars in binaries always rotate along with their orbital motion. The stellar rotation will result in each nonradial oscillation with the spherical harmonic index $\ell$ splitting into $2\ell+1$ different frequencies, thus effects of rotation on oscillations should be included. In this work, we extend the work of S$\acute{\rm a}$nchez Arias et al. (2017), and perform a more comprehensive asteroseismic analysis for CoRoT 100866999.  The details of input physics are presented in Section 2, and model grids are elaborated in Section 3. We introduce our fitting results in Section 4 and discuss them in Section 5. Finally, we conclude the main results of this work in Section 6. 

\section{Input Physics}
The stellar evolution code Modules for Experiments in Stellar Astrophysics (MESA), which was developed by Paxton et al. (2011, 2013, 2015, 2018), is used to compute evolutionary and pulsational models. In particular, the submodule called "pulse\_adipls" in version 10398 is used to generate stellar evolutionary models (Paxton et al. 2011, 2013, 2015, 2018), and calculate adiabatic frequencies of their radial and nonradial modes (Christensen-Dalsgaard 2008).

In our work, the 2005 update of the OPAL equation of state tables (Rogers $\&$ Nayfonov 2002) are adopted. The OPAL opacity tables of Iglesias $\&$ Rogers (1996) for high temperatures region and tables of Ferguson et al. (2005) for low temperatures region are used. The initial ingredient in metallicity is assumed to be identical to that of the sun (Asplund et al. 2009). The classical mixing length theory of B$\ddot{\rm o}$hm-Vitense (1958) with $\alpha$ = 1.90 (Paxton et al. 2011) is used in the convective region. For the overshooting mixing of the convective core, we adopt an exponentially decaying prescription and introduce an overshooting mixing diffusion coefficient
\begin{equation}
D_{\rm ov}=D_0{\rm exp}(\frac{-2z}{f_{\rm ov}H_{\rm p}})
\end{equation}
(Freytag et al. 1996; Herwig 2000). In equation (1), $D_0$ is the diffusion mixing coefficient near the edge of the convective core, $z$ the distance into radiative zone away from the edge, $H_{\rm p}$ the pressure scale height, and $f_{\rm ov}$ an adjustable parameters describing the efficiency of the overshooting mixing. In our calculations, the lower limit of the diffusion coefficient is set to be $D_{\rm ov}^{\rm limit}$ = 1$\times$10$^{-2}$ cm$^2$ s$^{-1}$, below which overshooting shuts off. In addition, effects of the stellar rotation, the element diffusion, and magnetic fields on the stellar structure and evolution are not included in this work.
\section{Gird of stellar models}
The evolutionary track and the interior structure of a star depend on the initial stellar mass $M$, initial chemical ingredient ($X$, $Y$, and $Z$), and the overshooting parameter $f_{\rm ov}$. In our work, the initial helium abundance is set to be $Y = 0.249 + 1.33Z$ (Li et al. 2018),  as a function of the metallicity $Z$, thus the stellar structure and evolution can be completely determined by $M$, $Z$ and $f_{\rm ov}$. In our work, we consider stellar masses $M$ between 1.50 $M_{\odot}$ and 2.20 $M_{\odot}$ with a step of 0.01 $M_{\odot}$, and metallicities $Z$ between 0.005 to 0.030 with a step of 0.001. These values of $Z$ corresponds to the range of [Fe/H] from -0.4 to 0.4 according to
\begin{equation}
[{\rm Fe/H}]= \log(\frac{Z}{X})-\log(\frac{Z}{X})_{\odot},
\end{equation}
where we adopt the value of $(Z/X)_{\odot}$ = 0.0181 (Asplund et al. 2009). For the overshooting mixing, we also adopt four different cases: no overshooting ($f_{\rm ov}$ = 0), moderate overshooting ($f_{\rm ov} =0.01$), intermediate overshooting ($f_{\rm ov} =0.02$), and extreme overshooting ($f_{\rm ov} = 0.03$).

Each star in the grid is computed from the zero-age main sequence to the post-main sequence stage. Figure 1 depicts a set of evolutionary tracks of stars with $Z=0.012$, $f_{\rm ov} = 0.02$, and $M$ ranging from 1.50 $M_{\odot}$ to 2.20 $M_{\odot}$ in a interval of 0.01 $M_{\odot}$. The two dotted lines in Figure 1 corresponds to the constraint of the effective temperature 7000 K $< T_{\rm eff}<$ 8400 K (Chapellier $\&$ Mathias 2013; Sarro et al. 2013). For theoretical models meeting with the constraint, we calculate its frequencies of radial oscillations ($\ell = 0$) and nonradial oscillations with $\ell=1$ and $\ell = 2$.

Besides, following works of Chen $\&$ Li (2018, 2019), we consider the rotation period $P_{\rm rot}$ between 0 day and 10 days with a step of 0.1 days, as the fourth adjustable parameter. For a given $P_{\rm rot}$, each nonradial oscillation mode will split into $2\ell+1$ different frequencies according to 
\begin{equation}
\nu_{\ell,n,m}= \nu_{\ell,n} + m\delta\nu_{\ell,n} =  \nu_{\ell,n} + \beta_{\ell, n}\frac{m}{P_{\rm rot}}
\end{equation}
(Saio 1981; Dziembowski \& Goode 1992; and Aerts et al. 2010), where $\delta\nu_{\ell,n}$ is the splitting frequency, $R$ is the present radius of the star, and $\beta_{\ell,n}$ is the rotational parameter that determines the size of rotational splitting. 
The general expression of $\beta_{\ell,n}$ for a uniformly rotating star is deduced to be 
\begin{equation}
\beta_{\ell, n}=\frac{\int_{0}^{R}(\xi_{r}^{2}+L^{2}\xi_{h}^{2}-2\xi_{r}\xi_{h}-\xi_{h}^{2})r^{2}\rho dr}
{\int_{0}^{R}(\xi_{r}^{2}+L^{2}\xi_{h}^{2})r^{2}\rho dr}
\end{equation}
(Aerts et al. 2010), where $\xi_{r}$ and $\xi_{h}$ are the radial displacement and the horizontal displacement respectively, $\rho$ is the local density, and $L^{2}= \ell(\ell+1)$. According to equation (3), each oscillation mode of $\ell = 1$ splits into three different frequencies,  forming a triplet. Each oscillation mode of $\ell = 2$ splits into five different frequencies, forming a quintuplet. 

\section{Fitting Results of CoRoT 100866999}
Table 1 lists the eight independent $\delta$ Scuti frequencies obtained by Chapellier $\&$ Mathias (2013). The frequency $F$ is the largest-amplitude mode, almost 20 times larger than those of other $\delta$ Scuti frequencies. Besides, the period ratio of $p_2$ and $F$ is 0.776, which equals the well known period ratio 0.772 between the fundamental radial mode and the radial first overtone mode (Fitch 1981; Poretti et al. 2005). In general, the component stars in eclipsing binaries always rotate along with their orbital motion. Su{\'a}rez et al. (2006) investigated effects of rotation on period ratios of radial modes, and found that the difference of period ratios remains around $10^{-3}$ for rotational velocities up to 50 km s$^{-1}$. Moreover, Chapellier $\&$ Mathias (2013) found that the radial fundamental mode $F$ corresponds to an absolute magnitude $M_{\upsilon}$ = 2.4 mag and an A8V spectral type for the primary star according to the period-luminosity relation of Templeton et al. (2002). These values match well the parameters obtained from the eclipsing light curve fit. Therefore, we use $F$ as the fundamental radial mode and $p_2$ as the radial first overtone mode in our calculations.

To find the optimal model to observations, we compare model frequencies with the observed frequencies $F$, $p_{1}$, $p_{2}$, $p_{3}$,  $p_{4}$,  $p_{5}$, $p_{6}$, and $p_7$ according to 
\begin{equation}
S^{2}=\frac{1}{k}\sum(|\nu_{\rm mod,i}-\nu_{\rm obs, i}|^{2}),
\end{equation}
where $\nu_{\rm obs, i}$ represents the observed frequency, $\nu_{\rm mod, i}$ represents its corresponding model frequency, and $k$ is the number of the observed frequencies. Frequencies $p_{1}$, $p_{3}$, $p_{4}$,  $p_{5}$,  $p_{6}$, and $p_{7}$ are not identified in advanced, thus theoretical frequencies nearest to them is regarded as their possible model counterparts.

Figures 2 and 3 show plots of resulting $S_{\rm m}^{2}$ versus various physical parameters. Each circle in the figures represents one minimum value of $S^{2}$ along one evolutionary track. We denote the minimum value with $S_{\rm m}^{2}$. In the figures, circles in black, bule, red, and green correspond to theoretical models with $f_{\rm ov}$ = 0, 0.01, 0.02, and 0.03, respectively. The horizontal line in orange marks the position of $S_{\rm m}^2$ = 0.1, which corresponds to the square of four times of the frequency resolution $1/\Delta T$. The circles above the horizontal line corresponds to 78 candidate models in Table 2. The filled circle in the figures corresponds to the best-fitting model (Model 68).

Figure 2(a) presents changes of $S_{\rm m}^2$ as a function of the stellar mass $M$. In the figure, values of $M$ are found to cover a wide range, i.e., 1.67$-$1.84 $M_{\odot}$, while stellar masses of theoretical models with the same overshooting mixing are found to exhibit good convergence. Therein, values of $M$ converge well to 1.79$-$1.84 $M_{\odot}$ for models with $f_{\rm ov}=0$, to 1.75$-$1.77 $M_{\odot}$ for models with $f_{\rm ov} =0.01$, and to 1.67$-$1.71 $M_{\odot}$ for models with $f_{\rm ov} = 0.02$.

Figures 2(b) and 2(c) present changes of $S_{\rm m}^{2}$ as a function of the metallicity $Z$ and the rotation period $P_{\rm rot}$, respectively. It can be clearly seen in the figures that values of $Z$ converge well to 0.012$-$0.016, and those of $P_{\rm rot}$ converge well to 3.6$-$4.7 days.

Figure 2(d) presents changes of $S_{\rm m}^2$ as a function of the overshooting parameter $f_{\rm ov}$. As shown in the figure, the best-fitting model has an intermediate overshooting ($f_{\rm ov} =0.02$). Besides, we find that theoretical models without overshooting ($f_{\rm ov}$ = 0) and with moderate overshooting ($f_{\rm ov}$ = 0.01) can also reproduce well the eight $\delta$ Scuti frequencies.

Figures 3(a)-(d) depict changes of $S_{\rm m}^{2}$ as a function of the effective temperature $T_{\rm eff}$, the luminosity $L$, the stellar radial $R$, and the gravitational acceleration $\log g$, respectively. In Figures 3(a) and 3(b), it can be noticed that both of $T_{\rm eff}$ and $L$ distribute over a wide range, i.e., $T_{\rm eff}$ = 7727$-$8273 K and $L$ = 10.051$-$14.054 $L_{\odot}$. In Figures 3(c) and 3(d), $R$ and $\log g$ distribute over a relatively smaller range, $R$ = 1.771$-$1.827 $R_{\odot}$ and $\log g$ = 4.164$-$4.179. Moreover, $R$ and $\log g$ of theoretical models with the same overshooting are found to be in good convergence.

Figure 3(e) depicts changes of $S_{\rm m}^{2}$ as a function of the mass fraction of central hydrogen $X_{\rm c}$, and
Figure 3(f) depicts changes of $S_{\rm m}^{2}$ as a function of ages of stars. Figure 3(e) shows that $X_{\rm c}$ distribute between 0.468 and 0.499, and Figure 3(f) shows that their ages range from 0.567 Gyr to 0.891 Gyr. This results indicate that the primary star of CoRoT 100866999 is being in main sequence stage.

Based on the above considerations, fundamental parameters of the primary star derived from the asteroseismic models are listed in Table 3. They match well the parameters obtained from the eclipsing light curve fit given by Chapellier $\&$ Mathias (2013). Table 4 lists model frequencies of the best-fitting model. Table 5 lists comparisons between model frequencies of the best-fitting model and the observed $\delta$ Scuti frequencies. Based on the comparisons, frequencies $p_1$ and $p_5$ are identified as two dipole modes, and $p_3$, $p_4$, $p_6$, and $p_7$ as four quadrupole modes. In particular, frequencies $p_4$ and $p_7$ are identified as two components of one quintuplet.

\section{Discussions}
In Section 4, we have introduced our fitting results. Physical parameters of theoretical models with different overshooting have a certain dispersion, while those of theoretical models with the same overshooting exhibit good convergence. In order to explain this, propagating properties of the oscillation modes in the star are examined in detail.

Figure 4 illustrates the profiles of Brunt$-$V$\ddot{\rm a}$is$\ddot{\rm a}$l$\ddot{\rm a}$ frequency $N$, characteristic acoustic frequencies $S_{\ell}$ ($\ell$ = 1 and 2) and hydrogen abundance $X_{\rm H}$ inside the best-fitting model. Figure 5 illustrates the scaled eigenfunctions inside the best-fitting model. The vertical line in Figure 5 denotes the boundary of the convective core ($\nabla_{\rm r}=\nabla_{\rm ad}$). The inner zone is the convective core, and the outer zone is the radiation envelope. It can be clearly seen in Figure 5 that the fundamental radial mode $F$ and the radial first overtone mode $p_2$ propagate mainly in the stellar envelope, and then characterize the features of the stellar envelope. For those nonradial oscillation modes, we find that $p_1$ is predominantly a g mode with largest amplitude near the edge of the convective core, and the others exhibit a mixed character with substantial amplitudes near the edge of the convective core and the surface. Therefore, the nonradial oscillation modes can give strong constraints on conditions of the convective core .

The acoustic radius $\tau_0$ is the sound travel time between the center and the surface of the star. It is defined by Aerts et al. (2010) as 
\begin{equation}
\tau_0=\int_0^R\frac{dr}{c_s},
\end{equation}
where $c_{\rm s}$ the adiabatic sound speed. In general, the value of $c_s$ in the envelope is much smaller than that in the convective core, thus $\tau_0$ is suitable to characterize features of the stellar envelope. Given that the nonradial modes have substantial amplitudes near the edge of the convective core, we use the relative radius of the convective core $R_{\rm conv}/R$ to characterize features of the deep interior of the star.

To fit the eight $\delta$ Scuti frequencies, both the convective core and the stellar envelope of the theoretical model need to be matched to the actual structure of CoRoT 100866999. Figures 6 and 7 show changes of $S_{\rm m}^{2}$ as a function of $\tau_0$ and $R_{\rm conv}/R$, respectively. In the figures, we find that $\tau_0$ of the candidate models converge well to $7154^{+18}_{-61}$ s and $R_{\rm conv}/R$ of the candidate models converge well to $0.0931^{+0.0003}_{-0.0013}$. This suggests that they are nearly alike in structure. 

The relation between the large frequency separation and the mean density of $\delta$ Scuti stars have been investigated in details (e.g., Su\'arez et al. 2014; Garc\'ia Hern\'andez 2015). Garc\'ia Hern\'andez (2017) updates the relation as  
\begin{equation}
\bar{\rho}/\rho_{\odot} = 1.50^{+0.09}_{-0.10}(\Delta\nu/\Delta\nu_{\odot})^{2.04^{+0.04}_{-0.04}},
\end{equation}
where $\Delta\nu_{\odot}$ = 134.8 $\mu$Hz (Kjeldsen, Bedding \& Christensen- Dalsgaard 2008). 
For the best-fitting model, the mean density $\bar{\rho}_{\rm mod}$ and the surface gravity $\log g_{\rm mod}$ are estimated to be 0.422 g/cm$^3$ and 4.166, respectively. Besides, we estimate the averaged frequency spacing $(\Delta\nu)_{\rm avg}$ of p modes in Table 4 to be 59.258 $\mu$Hz based on the best-fitting model. According to equation (7), $(\Delta\nu)_{\rm avg}$ corresponds to a mean density of 0.395 $\pm$ 0.038 g/cm$^3$ and a surface gravity of 4.138$^{+0.040}_{-0.044}$, matching well the mean density and surface gravity of the best-fitting model.

The rotation period $P_{\rm rot}$ of the primary star is determined to be $P_{\rm rot}=4.1^{+0.6}_{-0.5}$ days, which is slower than the orbital period $P_{\rm orb}$ = 2.80889 days. Ouazzani et al. (2010) investigated the rotational splittings of $\beta$ Cephei stars, and found that the threshold of validity of perturbative methods is extended to 10\% of the break-up velocity. The break-up velocity $\upsilon_{\rm crit}$ of the primary star is estimated to be 424 $\pm$ 2 km s$^{-1}$. The rotation velocity $\upsilon_{\rm e}$ is estimated to be $22\pm3$  km s$^{-1}$ according to $\upsilon_{\rm e}$=$2\pi R/P_{\rm rot}$, about 5.2\% of the break-up velocity $\upsilon_{\rm crit}$, thus the linear perturbation method is still valid. Moreover, given that its low rotational velocity, the effect of rotation on the stellar structure and evolution are not included in our calculations. According to work of Saio (1981), Dziembowski \& Goode (1992), and Aerts et al. (2010), the first-order effect of rotation on pulsation is in proportion to $1/P_{\rm rot}$, and that of the second-order is in proportion to $1/(P_{\rm rot}^2\nu_{\ell, n})$. The ratio of the second-order and the first-order effect of rotation can be estimated to be in the order of $1/(P_{\rm rot}\nu_{\ell, n})$, where 1/$P_{\rm rot}$ is estimated to be $2.82^{+0.39}_{-0.36}$ $\mu$Hz and $\nu_{\ell, n}$ ranges from 188.113 $\mu$Hz to 308.669 $\mu$Hz. The second-order effect of rotation is much less than that of the first-order one . Hence the second-order effect of rotation on pulsation is not included in this work.

Gizon $\&$ Solanki (2003) investigated the relation of amplitude of the nonradial oscillation and the inclination angle $i$ of the stellar rotation axes in details. The inclination angle $i$ of CoRoT 100866999 is determined to be 80 $\pm$ 2 degrees (Chapellier $\&$ Mathias 2013). According to work of Gizon $\&$ Solanki (2003), dipole modes with $m$ = $\pm$ 1 are more visible than those with $m=0$, and quadrupole modes with $m$ = 0 and $\pm$ 2 exhibit more visible than those with $m$ = $\pm$ 1. As shown in Table 5, amplitudes of $p_1$, $p_3$, $p_4$, and $p_6$ meet the predictions, while those of $p_5$ and $p_7$ do not conform with the relation. The studies of Gizon $\&$ Solanki (2003) is on the basis of the assumption that energy equals each other between modes with different azimuthal number. However, this assumption is not always valid for $\delta$ Scuti stars, such as for KIC 9244992 (Saio et al. 2015), the relative amplitudes of modes with $m=0$ and those with $m$ = $\pm$ 1 varies in different multiplets.

Finally, it should be pointed out that all of the above analyses are based on the $\delta$ Scuti frequencies. The pulsating primary star is a hybrid $\delta$ Sct-$\gamma$ Dor star with two clearly distinct frequencies domains (Chapellier $\&$ Mathias 2013). Except for the independent $\delta$ Scuti frequencies, there are 63 independent $\gamma$ Doradus frequencies. In Table 4, it can be seen that model frequencies in the interval [3; 60] $\mu$Hz are very dense, including 94 theoretical frequencies with $\ell = 1$ and 165 theoretical frequencies with $\ell =2$. According to equation (3), each mode with $\ell = 1$ will split into three different frequencies, and each mode with $\ell = 2$ will split into five different frequencies. Considering the effects of rotation on pulsation, model frequencies will become much denser.

Based on the best-fitting model, we compare model frequencies with the 63 $\gamma$ Doradus frequencies. The model frequency nearest to observations is regarded as its possible model counterpart, and the comparing results are listed in Table 6. In Table 6, the 22 frequencies with a nearly constant period separation identified by Chapellier $\&$ Mathias (2013) are marked in boldface. It can be seen in Table 6 that these 22 frequencies can be well reproduced based on our best-fitting model, and their period differences between model frequencies of the best-fitting model and observations range from 0 days for $f_{38}$ to 0.0035 days for $f_{46}$. Deviations from the rigorously uniform period spacing for model frequencies nearly equal to those for the 22 observed modes. Furthermore, we compare the $\gamma$ Dor frequencies with other stellar models in the grid, due to very dense of the model frequencies, the $\gamma$ Dor frequencies can also be well reproduced. Therefore, we do not use the $\gamma$ Doradus frequencies to restrict our theoretical models in this work.

\section{Summary and Conclusions}
We have carried out largely numerical calculations and detailed asteroseismic analyses for the pulsating primary star of CoRoT 100866999. The  main results of this work are concluded as follows:

1. The pulsating primary star of CoRoT 100866999 is found to be a main sequence star with a rotation period of $4.1^{+0.6}_{-0.5}$ days. The primary star rotates slower than the orbital motion.

2. Physical parameters of the primary star of CoRoT 100866999 are determined to be $M$ = $1.71^{+0.13}_{-0.04}$ $M_{\odot}$, $Z=0.012^{+0.004}_{-0.000}$, $f_{\rm ov}$ = $0.02^{+0.00}_{-0.02}$, $T_{\rm eff}$ = $8024^{+249}_{-297}$ K,  $L$ = $11.898^{+2.156}_{-1.847}$ $L_{\odot}$, $\log g$ = $4.166^{+0.013}_{-0.002}$, $R$ = $1.787^{+0.040}_{-0.016}$ $R_{\odot}$, and $X_{\rm c}$ = 0.488$^{+0.011}_{-0.020}$. The asteroseismic parameters match well those from the eclipsing curve fit given by Chapellier $\&$ Mathias (2013).

3. Based on comparisons between model frequencies and observations, $p_1$ and $p_5$ are identified as two dipole modes, and $p_3$, $p_4$, $p_6$, and $p_7$ as four quadrupole modes. Moreover, we find that frequencies $p_4$ and $p_7$ are two component of one quintuplet.

4. Based on the best-fitting model, $p_1$ is found to be a g mode, and the other nonradial modes are found to exhibit distinct mixed characters. These modes provide strong constraints on conditions of the deep interior of the star. Finally, the relative radius of the convective core of CoRoT 100866999 is suggested to be $R_{\rm conv}/R$ = $0.0931^{+0.0003}_{-0.0013}$.
\acknowledgments
The authors thank for the fund from the NSFC of China (Grant No. 11333006, 11521303, 11833002, 11803082). Xinghao Chen appreciate the support the West Light Foundation of The Chinese Academy of Sciences. The authors gratefully acknowledge the computing time granted by the Yunnan Observatories, and provided on the facilities at the Yunnan Observatories Supercomputing Platform. The authors are sincerely grateful to an anonymous referee for instructive advice and productive suggestions. Xinghao Chen is also thankful for fruitful discussions with Jianheng Guo, Jie Su, and Tao Wu.

\appendix 
\section{Inlist file of pulse\_adipls in MESA (Version 10398)}
\begin{verbatim} 
! inlist_pulse
&star_job
 astero_just_call_my_extras_check_model = .true.
 create_pre_main_sequence_model = .true.
 change_lnPgas_flag = .true.
 new_lnPgas_flag = .true.
 change_initial_net = .true.
 new_net_name = 'o18_and_ne22.net'
 kappa_file_prefix = 'a09'
 kappa_lowT_prefix = 'lowT_fa05_a09p'
 initial_zfracs = 6
 relax_to_this_tau_factor = 1.5d-3
 relax_tau_factor = .true.
/ ! end of star_job namelist
&controls
 initial_mass = 1.71
 initial_z =    0.012
 initial_y =    0.26496
 overshoot_f_above_burn_h_core = 0.020

 min_overshoot_q = 1d-3
 D_mix_ov_limit = 1d-2
 overshoot_f0_above_burn_h_core = 0.005
 MLT_option = 'ML1'  
 mixing_length_alpha = 1.90
 calculate_Brunt_N2 =.true.
 use_brunt_gradmuX_form = .true.
 which_atm_option = 'simple_photosphere' !default

 history_interval = 1
 max_num_profile_models = 80000
 xa_central_lower_limit_species(1) = 'h1'
 xa_central_lower_limit(1) = 5d-5

 use_other_mesh_functions= .true.
 mesh_delta_coeff = 0.90
 M_function_weight = 50
 max_center_cell_dq = 1d-10
 varcontrol_target = 5d-5 !for main-sequence models (2d-4 for pre-main sequence models)
 max_years_for_timestep = 1d6
/ ! end of controls namelist
\end{verbatim}

\begin{figure*}
\includegraphics[width=0.8\textwidth, angle = 0]{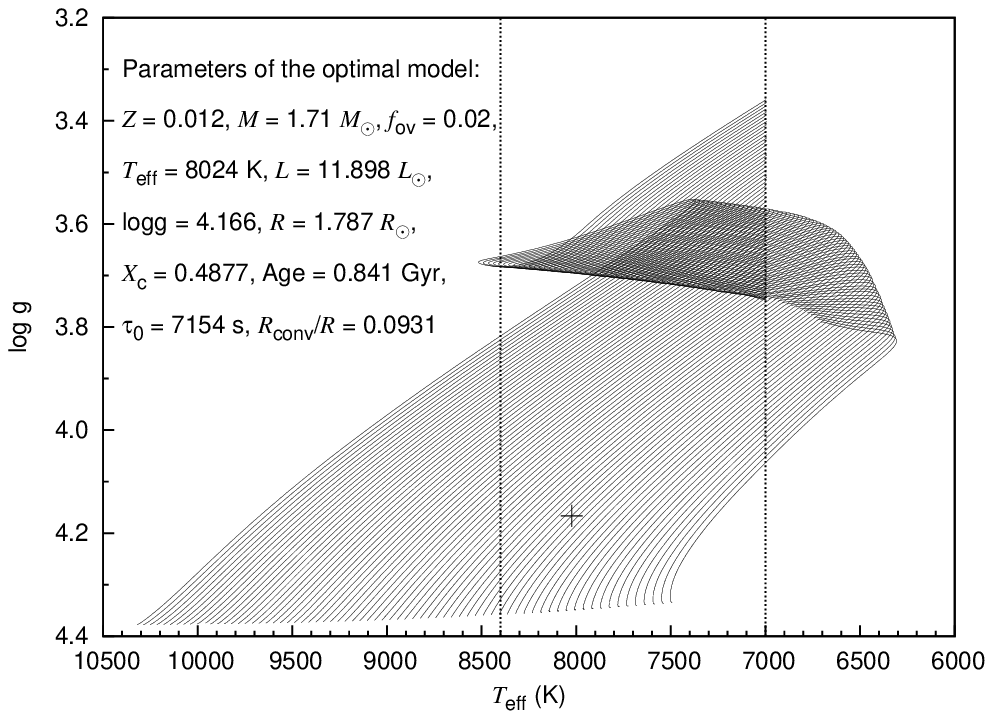}
  \caption{\label{Figure 1} Evolutionary tracks of  stars with $Z$ = 0.012, $f_{\rm ov}$ = 0.02, and $M$ ranging from 1.50 to 2.20 $M_{\odot}$ in a step of 0.01 $M_{\odot}$. The two dotted lines mark the constraints 7000 K $<$ $T_{\rm eff}$ $<$ 8400 K. The crossing denotes the best-fitting model, whose fundamental parameters are labelled in the top of the figure.}
\end{figure*}
\begin{figure*}
\includegraphics[width=0.8\textwidth, angle =0]{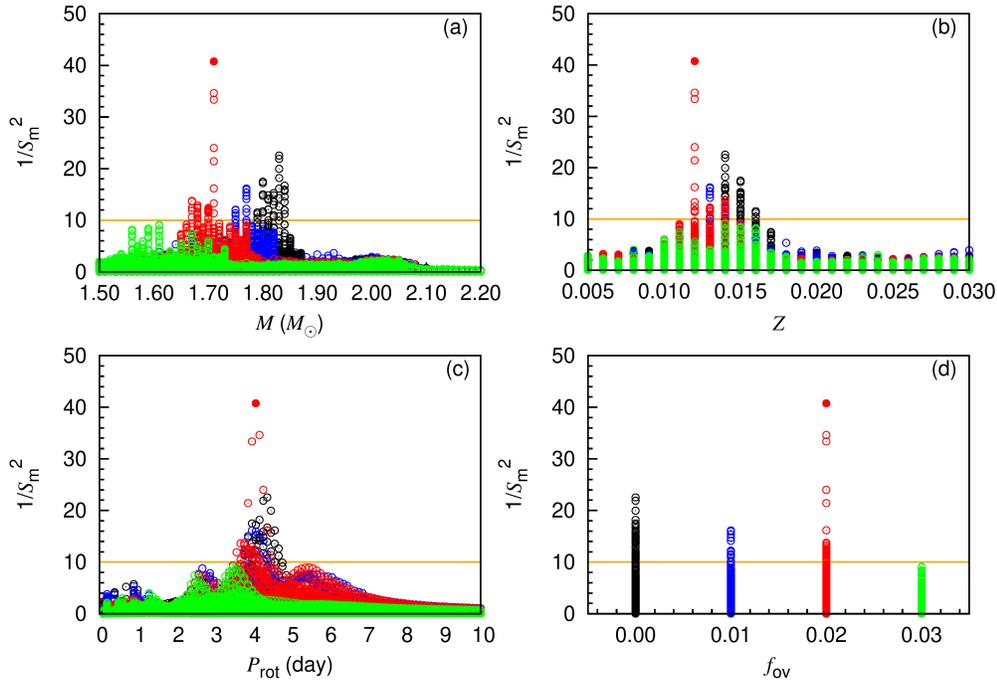}
\caption{\label{Figure 2} Visualisation of fitting results 1/$S_{\rm m}^{2}$ versus adjustable parameters: stellar mass $M$, the initial metallicity $Z$, the rotation period $P_{\rm rot}$, and the overshooting parameter $f_{\rm ov}$, respectively. The circles in black, blue, red, and green corresponds to stellar models with $f_{\rm ov}$ = 0, 0.01, 0.02, and 0.03, respectively. The horizontal line in orange mark the position of $S_{\rm m}^2$ = 0.1.}
\end{figure*}
\begin{figure*}
\includegraphics[width=0.8\textwidth, angle = 0]{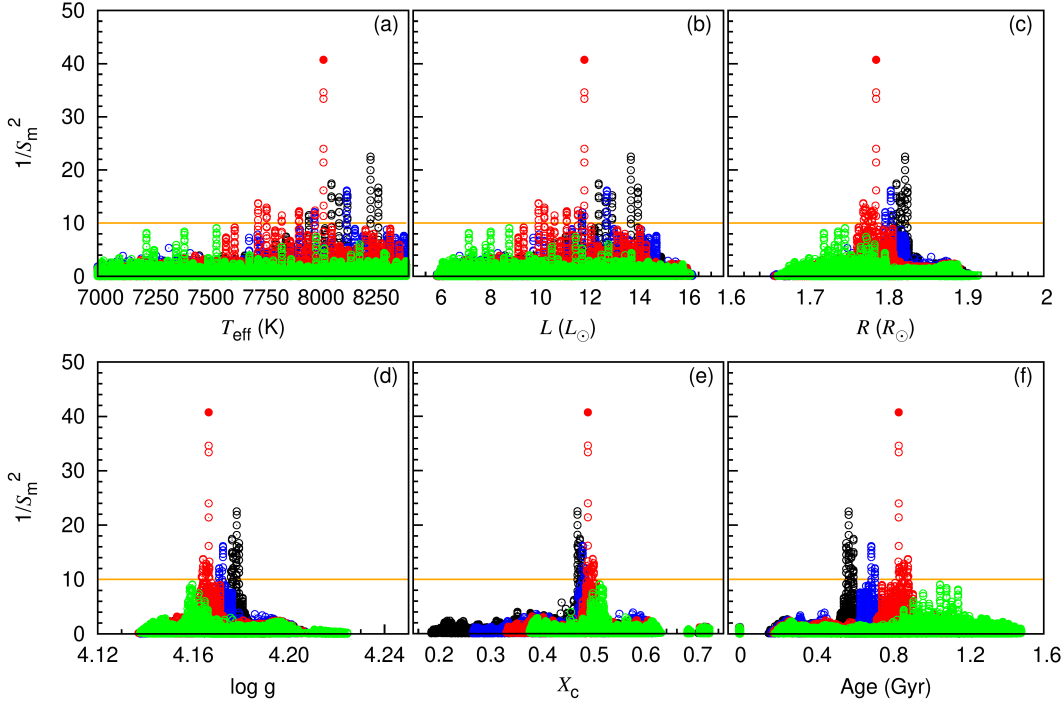}
\caption{\label{Figure 3} Visualisation of fitting results 1/$S_{\rm m}^{2}$ versus stellar fundamental parameters: the effective temperature $T_{\rm eff}$, the luminosity $L$, stellar radius R, the gravitational acceleration $\log g$, the mass fraction of central hydrogen $X_{\rm c}$, and the age of stars, respectively. The circles in black, blue, red, and green corresponds to stellar models with $f_{\rm ov}$ = 0, 0.01, 0.02, and 0.03, respectively. The horizontal line in orange mark the position of $S_{\rm m}^2$ = 0.1.}
\end{figure*}
\begin{figure*}
\includegraphics[width=0.8\textwidth, angle = 0]{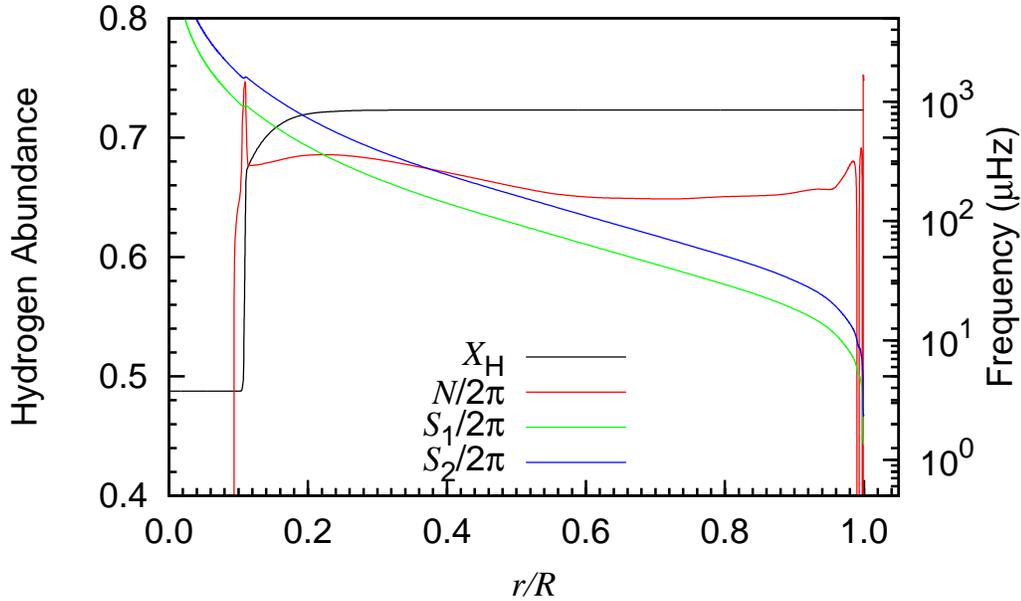}
\caption{\label{Figure 4} Brunt$-$V$\ddot{\rm a}$is$\ddot{\rm a}$l$\ddot{\rm a}$ frequency $N$, characteristic acoustic frequencies $S_{\ell}$ ($\ell$ = 1 and 2), and hydrogen abundance $X_{\rm H}$ inside the best-fitting model of CoRoT 100866999.}
\end{figure*}
\begin{figure*}
\includegraphics[width=0.8\textwidth, angle = 0]{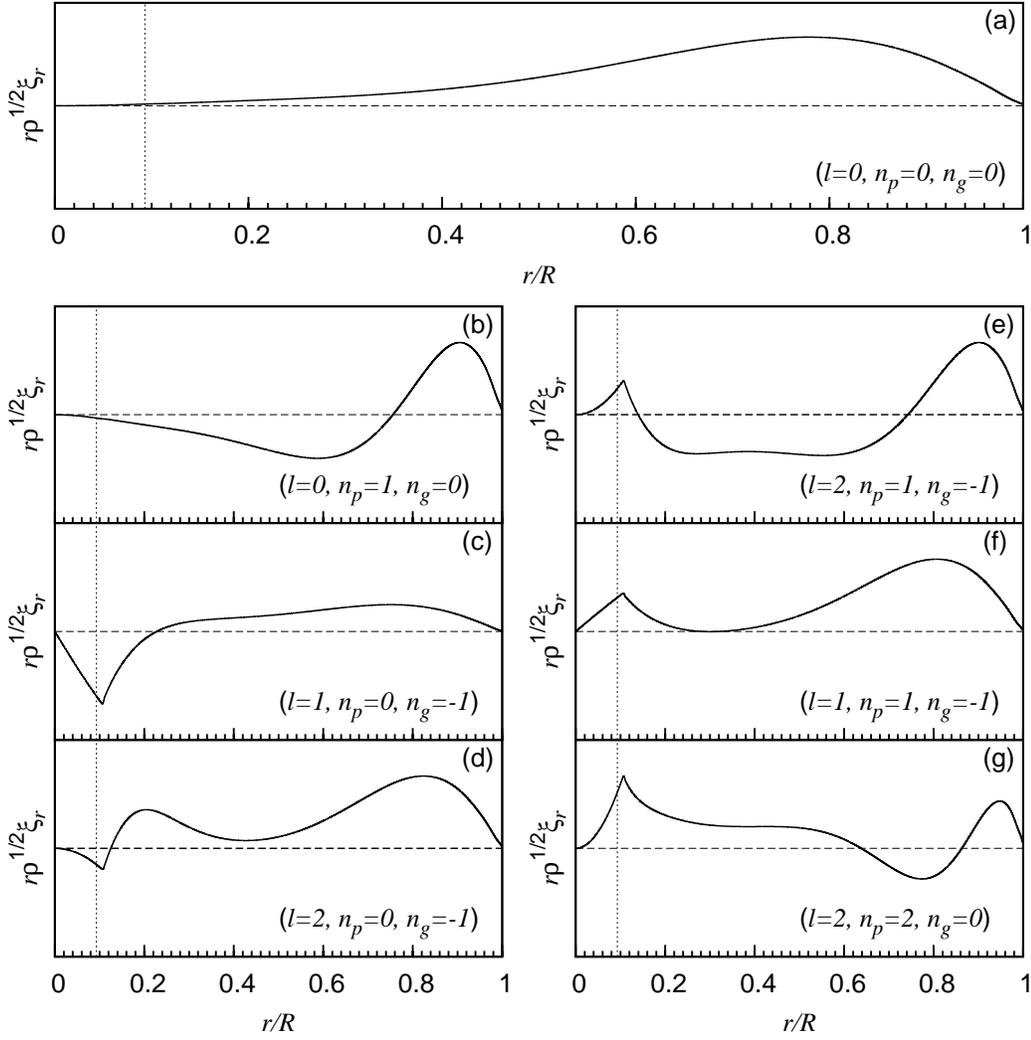}
\caption{\label{Figure 5} Scaled radial displacement eigenfunctions for theoretical modes corresponding to the observed frequencies inside the best-fitting model. Panel (a) corresponds to the fundamental radial mode $F$. Panel (b) corresponds to the radial first overtone mode $p_2$. Panels (c), (d), (f), and (g) correspond to the observed frequencies $p_1$, $p_3$, $p_5$, and $p_6$, respectively. Panel (e) corresponds to the observed frequencies $p_4$ and $p_7$. Vertical line marks the boundary of the convective core ($\nabla_{\rm r}=\nabla_{\rm ad}$).}
\end{figure*}
\begin{figure*}
\includegraphics[width=1.5\textwidth, angle = 0]{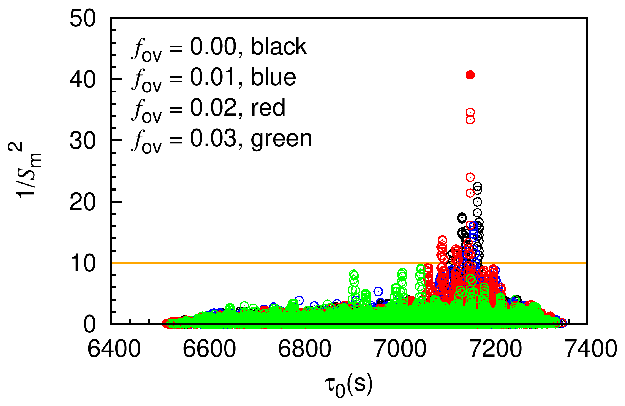}
\caption{\label{Figure 6} Visualisation of fitting results 1/$S_{\rm m}^{2}$ versus the acoustic radius $\tau_0$.
he circles in black, blue, red, and green corresponds to stellar models with $f_{\rm ov}$ = 0, 0.01, 0.02, and 0.03, respectively. The horizontal line in orange mark the position of $S_{\rm m}^2$ = 0.1.}
\end{figure*}
\begin{figure*}
\includegraphics[width=1.5\textwidth, angle = 0]{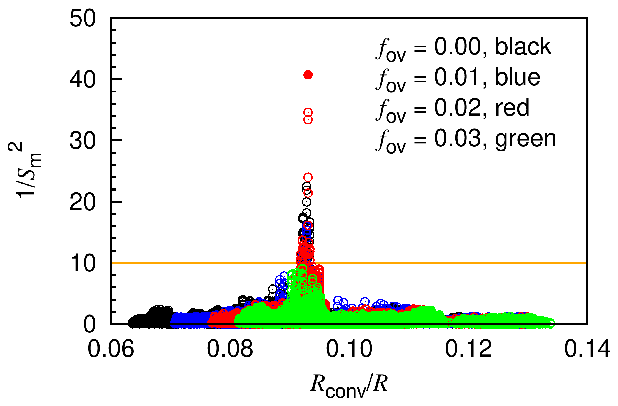}
\caption{\label{Figure 7} Visualisation of fitting results 1/$S_{\rm m}^{2}$ versus the relative radius of the convective core $R_{\rm conv}/R$. The circles in black, blue, red, and green corresponds to stellar models with $f_{\rm ov}$ = 0, 0.01, 0.02, and 0.03, respectively. The horizontal line in orange mark the position of $S_{\rm m}^2$ = 0.1.}
\end{figure*}
\begin{table*}
\centering
\caption{\label{t1}Eight independent $\delta$ Scuti frequencies obtained by Chapellier $\&$ Mathias (2013). ID represents the serial number of the observed frequencies in Chapellier $\&$ Mathias (2013). Freq. represents the observed frequency. Ampl. represents the amplitudes in unit of mmag. S/N represents the signal-to-noise ratio.}
\begin{tabular}{lccccc}
\hline\hline
ID                  &Freq.             &Freq.              &Ampl.       & S/N\\
                     &(d$^{-1}$)        &($\mu$Hz)          &(mmag)    &  \\
\hline
$F$               &16.9803            &196.531          &11.623      &515.5\\
$p_{1}$         &16.2530            &188.113          &0.508       &22.8\\
$p_{2}$         &21.8711            &253.138          &0.449       &23.3\\
$p_{3}$         &17.5521            &203.149          &0.223       &10.1\\
$p_{4}$         &21.6053            &250.061          &0.167       &8.5\\
$p_{5}$         &17.5674            &203.326          &0.151       &6.8\\
$p_{6}$         &26.6690            &308.669          &0.126       &6.6\\
$p_{7}$         &21.8441            &252.825          &0.125       &6.5\\
\hline
\end{tabular}
\end{table*}
\begin{table*}
\tiny
\centering
\caption{\label{t2}Candidate models with $S_{\rm m}^{2}$ $<$ 0.10. The first column is the serial number of candidate models. $P_{\rm rot}$ is the rotation period of the star. $\tau_0$ is the acoustic radius defined as equation (6). $R_{\rm conv}/R$ is the relative radius of the convective core in the star. $X_{\rm c}$ is the mass fraction of hydrogen in the center of the star. }
\begin{tabular}{ccccccccccccccccc}
\hline\hline
Model &$P_{\rm rot}$ &$Z$  &$M$   &$f_{\rm ov}$ &$T_{\rm eff}$ &$L$    &log$g$ &$R$ &$\tau_0$ &$R_{\rm conv}/R$ &$X_{\rm c}$ &Age &$S_{\rm m}^{2}$  \\
          &(day)  &     &($M_{\odot}$) &   &(K)  &$(L_{\odot})$  &(dex) &($R_{\odot}$) &(s)  &(1d-2)   &&(Gyr)  &  \\
\hline
 1 &3.80 &0.015 &1.80 &0.00 &8061 &12.472 &4.176 &1.813 &7137 &9.224 &0.4723 &0.605 &0.0875\\
 2 &3.80 &0.016 &1.79 &0.00 &7959 &11.788 &4.176 &1.809 &7113 &9.200 &0.4806 &0.604 &0.0890\\
 3 &3.90 &0.015 &1.80 &0.00 &8061 &12.472 &4.176 &1.813 &7137 &9.224 &0.4723 &0.605 &0.0665\\
 4 &3.90 &0.015 &1.82 &0.00 &8133 &13.015 &4.178 &1.820 &7145 &9.297 &0.4758 &0.580 &0.0910\\
 5 &3.90 &0.016 &1.79 &0.00 &7957 &11.791 &4.176 &1.809 &7117 &9.196 &0.4802 &0.605 &0.0868\\
 6 &4.00 &0.014 &1.83 &0.00 &8238 &13.767 &4.178 &1.824 &7169 &9.284 &0.4677 &0.580 &0.0930\\
 7 &4.00 &0.015 &1.80 &0.00 &8061 &12.472 &4.176 &1.813 &7136 &9.224 &0.4723 &0.605 &0.0572\\
 8 &4.00 &0.015 &1.82 &0.00 &8133 &13.015 &4.178 &1.820 &7145 &9.297 &0.4758 &0.580 &0.0730\\
 9 &4.00 &0.016 &1.79 &0.00 &7957 &11.791 &4.176 &1.809 &7117 &9.196 &0.4802 &0.605 &0.0957\\
10 &4.10 &0.014 &1.83 &0.00 &8238 &13.767 &4.178 &1.824 &7169 &9.284 &0.4677 &0.580 &0.0699\\
11 &4.10 &0.014 &1.84 &0.00 &8273 &14.054 &4.179 &1.827 &7172 &9.333 &0.4691 &0.568 &0.0836\\
12 &4.10 &0.015 &1.80 &0.00 &8061 &12.472 &4.176 &1.813 &7137 &9.224 &0.4723 &0.605 &0.0582\\
13 &4.10 &0.015 &1.81 &0.00 &8095 &12.746 &4.177 &1.817 &7146 &9.255 &0.4734 &0.594 &0.0826\\
14 &4.10 &0.015 &1.82 &0.00 &8133 &13.015 &4.178 &1.820 &7145 &9.297 &0.4758 &0.580 &0.0655\\
15 &4.20 &0.014 &1.83 &0.00 &8238 &13.767 &4.178 &1.824 &7169 &9.284 &0.4677 &0.580 &0.0550\\
16 &4.20 &0.014 &1.84 &0.00 &8273 &14.054 &4.179 &1.827 &7172 &9.333 &0.4691 &0.568 &0.0687\\
17 &4.20 &0.015 &1.80 &0.00 &8061 &12.472 &4.176 &1.813 &7134 &9.224 &0.4723 &0.605 &0.0679\\
18 &4.20 &0.015 &1.81 &0.00 &8095 &12.746 &4.177 &1.817 &7146 &9.255 &0.4734 &0.594 &0.0705\\
19 &4.20 &0.015 &1.82 &0.00 &8133 &13.015 &4.178 &1.820 &7145 &9.297 &0.4758 &0.580 &0.0673\\
20 &4.30 &0.014 &1.83 &0.00 &8238 &13.767 &4.178 &1.824 &7169 &9.284 &0.4677 &0.580 &0.0457\\
21 &4.30 &0.014 &1.84 &0.00 &8273 &14.054 &4.179 &1.827 &7172 &9.333 &0.4691 &0.568 &0.0629\\
22 &4.30 &0.015 &1.80 &0.00 &8061 &12.472 &4.176 &1.813 &7137 &9.224 &0.4723 &0.605 &0.0852\\
23 &4.30 &0.015 &1.81 &0.00 &8095 &12.746 &4.177 &1.817 &7146 &9.255 &0.4734 &0.594 &0.0673\\
24 &4.30 &0.015 &1.82 &0.00 &8133 &13.015 &4.178 &1.820 &7145 &9.297 &0.4758 &0.580 &0.0770\\
25 &4.40 &0.014 &1.83 &0.00 &8238 &13.767 &4.178 &1.824 &7169 &9.284 &0.4677 &0.580 &0.0444\\
26 &4.40 &0.014 &1.84 &0.00 &8273 &14.054 &4.179 &1.827 &7172 &9.333 &0.4691 &0.568 &0.0599\\
27 &4.40 &0.015 &1.81 &0.00 &8095 &12.746 &4.177 &1.817 &7146 &9.255 &0.4734 &0.594 &0.0718\\
28 &4.40 &0.015 &1.82 &0.00 &8133 &13.015 &4.178 &1.820 &7145 &9.297 &0.4758 &0.580 &0.0935\\
29 &4.50 &0.014 &1.83 &0.00 &8238 &13.767 &4.178 &1.824 &7169 &9.284 &0.4677 &0.580 &0.0501\\
30 &4.50 &0.014 &1.84 &0.00 &8273 &14.054 &4.179 &1.827 &7172 &9.333 &0.4691 &0.568 &0.0638\\
31 &4.50 &0.015 &1.81 &0.00 &8095 &12.746 &4.177 &1.817 &7146 &9.255 &0.4734 &0.594 &0.0831\\
32 &4.60 &0.014 &1.83 &0.00 &8238 &13.767 &4.178 &1.824 &7169 &9.284 &0.4677 &0.580 &0.0620\\
33 &4.60 &0.014 &1.84 &0.00 &8273 &14.054 &4.179 &1.827 &7172 &9.333 &0.4691 &0.568 &0.0738\\
34 &4.70 &0.014 &1.83 &0.00 &8238 &13.767 &4.178 &1.824 &7169 &9.284 &0.4677 &0.580 &0.0791\\
35 &4.70 &0.014 &1.84 &0.00 &8273 &14.054 &4.179 &1.827 &7172 &9.333 &0.4691 &0.568 &0.0891\\
36 &3.80 &0.013 &1.77 &0.01 &8130 &12.802 &4.172 &1.806 &7160 &9.288 &0.4779 &0.698 &0.0817\\
37 &3.90 &0.013 &1.77 &0.01 &8130 &12.802 &4.172 &1.806 &7160 &9.288 &0.4779 &0.698 &0.0680\\
38 &3.90 &0.014 &1.75 &0.01 &7984 &11.814 &4.171 &1.799 &7138 &9.223 &0.4828 &0.717 &0.0910\\
39 &4.00 &0.013 &1.77 &0.01 &8130 &12.802 &4.172 &1.806 &7160 &9.288 &0.4779 &0.698 &0.0619\\
40 &4.00 &0.014 &1.75 &0.01 &7984 &11.814 &4.171 &1.799 &7138 &9.223 &0.4828 &0.717 &0.0836\\
41 &4.10 &0.013 &1.77 &0.01 &8130 &12.802 &4.172 &1.806 &7160 &9.288 &0.4779 &0.698 &0.0627\\
42 &4.10 &0.014 &1.75 &0.01 &7984 &11.814 &4.171 &1.799 &7138 &9.223 &0.4828 &0.717 &0.0829\\
43 &4.20 &0.013 &1.77 &0.01 &8130 &12.802 &4.172 &1.806 &7160 &9.288 &0.4779 &0.698 &0.0650\\
44 &4.20 &0.014 &1.75 &0.01 &7984 &11.814 &4.171 &1.799 &7138 &9.223 &0.4828 &0.717 &0.0879\\
45 &4.30 &0.013 &1.77 &0.01 &8130 &12.802 &4.172 &1.806 &7160 &9.288 &0.4779 &0.698 &0.0708\\
46 &4.30 &0.014 &1.75 &0.01 &7984 &11.814 &4.171 &1.799 &7138 &9.223 &0.4828 &0.717 &0.0978\\
47 &4.40 &0.013 &1.77 &0.01 &8129 &12.806 &4.172 &1.807 &7165 &9.283 &0.4774 &0.699 &0.0867\\
48 &4.50 &0.013 &1.77 &0.01 &8129 &12.806 &4.172 &1.807 &7165 &9.283 &0.4774 &0.699 &0.0999\\
49 &3.60 &0.014 &1.67 &0.02 &7728 &10.051 &4.164 &1.771 &7095 &9.213 &0.4971 &0.890 &0.0840\\
50 &3.70 &0.013 &1.70 &0.02 &7914 &11.185 &4.167 &1.782 &7124 &9.337 &0.4945 &0.845 &0.0955\\
51 &3.70 &0.014 &1.67 &0.02 &7728 &10.051 &4.164 &1.771 &7095 &9.213 &0.4971 &0.890 &0.0734\\
52 &3.80 &0.012 &1.70 &0.02 &7986 &11.633 &4.165 &1.784 &7150 &9.283 &0.4861 &0.860 &0.0857\\
53 &3.80 &0.012 &1.71 &0.02 &8025 &11.894 &4.167 &1.787 &7149 &9.313 &0.4880 &0.840 &0.0753\\
54 &3.80 &0.013 &1.70 &0.02 &7914 &11.185 &4.167 &1.782 &7124 &9.337 &0.4945 &0.845 &0.0840\\
55 &3.80 &0.014 &1.67 &0.02 &7728 &10.051 &4.164 &1.771 &7095 &9.213 &0.4971 &0.890 &0.0728\\
56 &3.80 &0.014 &1.68 &0.02 &7767 &10.282 &4.166 &1.773 &7093 &9.233 &0.4994 &0.868 &0.0890\\
57 &3.90 &0.012 &1.70 &0.02 &7986 &11.633 &4.165 &1.784 &7150 &9.283 &0.4861 &0.860 &0.0807\\
58 &3.90 &0.012 &1.71 &0.02 &8024 &11.898 &4.166 &1.787 &7154 &9.309 &0.4877 &0.841 &0.0467\\
59 &3.90 &0.013 &1.70 &0.02 &7914 &11.185 &4.167 &1.782 &7124 &9.337 &0.4945 &0.845 &0.0816\\
60 &3.90 &0.014 &1.67 &0.02 &7728 &10.051 &4.164 &1.771 &7095 &9.213 &0.4971 &0.890 &0.0806\\
61 &3.90 &0.014 &1.68 &0.02 &7767 &10.282 &4.166 &1.773 &7093 &9.233 &0.4994 &0.868 &0.0794\\
62 &4.00 &0.012 &1.70 &0.02 &7986 &11.633 &4.165 &1.784 &7150 &9.283 &0.4861 &0.860 &0.0876\\
63 &4.00 &0.012 &1.71 &0.02 &8024 &11.898 &4.166 &1.787 &7154 &9.309 &0.4877 &0.841 &0.0300\\
64 &4.00 &0.013 &1.68 &0.02 &7835 &10.692 &4.164 &1.777 &7126 &9.186 &0.4902 &0.886 &0.0974\\
65 &4.00 &0.013 &1.70 &0.02 &7913 &11.189 &4.166 &1.782 &7127 &9.334 &0.4942 &0.846 &0.0828\\
66 &4.00 &0.014 &1.67 &0.02 &7727 &10.054 &4.164 &1.772 &7099 &9.209 &0.4968 &0.891 &0.0944\\
67 &4.00 &0.014 &1.68 &0.02 &7766 &10.285 &4.165 &1.774 &7095 &9.229 &0.4990 &0.869 &0.0774\\
68 &4.10 &0.012 &1.71 &0.02 &8024 &11.898 &4.166 &1.787 &7154 &9.309 &0.4877 &0.841 &0.0246\\
69 &4.10 &0.013 &1.68 &0.02 &7835 &10.692 &4.164 &1.777 &7126 &9.186 &0.4902 &0.886 &0.0872\\
70 &4.10 &0.013 &1.70 &0.02 &7913 &11.189 &4.166 &1.782 &7127 &9.334 &0.4942 &0.846 &0.0869\\
71 &4.10 &0.014 &1.68 &0.02 &7766 &10.285 &4.165 &1.774 &7095 &9.229 &0.4990 &0.869 &0.0810\\
72 &4.20 &0.012 &1.71 &0.02 &8024 &11.898 &4.166 &1.787 &7154 &9.309 &0.4877 &0.841 &0.0289\\
73 &4.20 &0.013 &1.68 &0.02 &7834 &10.695 &4.164 &1.778 &7129 &9.182 &0.4899 &0.887 &0.0952\\
74 &4.20 &0.013 &1.70 &0.02 &7913 &11.189 &4.166 &1.782 &7127 &9.334 &0.4942 &0.846 &0.0971\\
75 &4.20 &0.014 &1.68 &0.02 &7766 &10.285 &4.165 &1.774 &7095 &9.229 &0.4990 &0.869 &0.0913\\
76 &4.30 &0.012 &1.71 &0.02 &8024 &11.898 &4.166 &1.787 &7154 &9.309 &0.4877 &0.841 &0.0417\\
77 &4.40 &0.012 &1.71 &0.02 &8024 &11.898 &4.166 &1.787 &7154 &9.309 &0.4877 &0.841 &0.0618\\
78 &4.50 &0.012 &1.71 &0.02 &8024 &11.898 &4.166 &1.787 &7154 &9.309 &0.4877 &0.841 &0.0882\\
\hline
\end{tabular}
\end{table*}
\begin{table*}
\centering
\caption{\label{t3} Fundamental parameters of the primary star of CoRoT 100866999. $P_{\rm rot}$ represents the rotation period of the star. $\tau_0$ represents the acoustic radius defined as equation (6). $R_{\rm conv}/R$ is the relative radius of the convective core in the star. $X_{\rm c}$ represents the mass fraction of hydrogen in the center of the star.}
\begin{tabular}{llll}
\hline\hline
Parameter                                                  &Values                         \\
\hline
$M(M_{\odot})$                                         &$1.67-1.84$ ($1.71^{+0.13}_{-0.04}$)\\
$Z$                                                           &$0.012-0.016$ ($0.012^{+0.004}_{-0.000}$)\\
$f_{\rm ov}$                                              &$0-0.02$ ($0.02_{-0.02}^{+0.00}$)\\
$P_{\rm rot}$(day)                                     &$3.6-4.7$ ($4.1^{+0.6}_{-0.5}$)\\
$T_{\rm eff}$ (K)                                        &$7727-8273$ ($8024^{+249}_{-297}$)\\
log$g$ (dex)                                              &$4.164-4.179$ ($4.166^{+0.013}_{-0.002}$)\\
$R$ $(R_{\odot})$                                      &$1.771-1.827$ ($1.787^{+0.040}_{-0.016}$)\\
$L$ $(L_{\odot})$                                       &$10.051-14.054$ ($11.898^{+2.156}_{-1.847}$)\\
$X_{\rm c}$                                                &$0.468-0.499$ ($0.488^{+0.011}_{-0.020}$)\\
Age (Gyr)                                                    &$0.567-0.891$ ($0.841^{+0.050}_{-0.274}$)\\
$\tau_0$ (s)                                                &$7093-7172$ ($7154^{+18}_{-61}$)\\
$R_{\rm conv}/R$                                       &$0.0918-0.0934$ ($0.0931^{+0.0003}_{-0.0013}$)\\
\hline
\end{tabular}
\end{table*}
\begin{table*}
\centering
\tiny
\caption{\label{t4}Theoretical model frequencies of the best-fitting model. $\nu_{\rm mod}$ is the model frequencies in unit of $\mu$Hz, $\ell$ is the spherical harmonic degree, $n_p$ and $n_g$ are the number of radial orders in p-mode propagation cavity and g-mode propagation cavity, respectively. $\beta_{\ell,n}$ is the rotational parameters defined as  Equation (3).}
\label{observed frequencies}
\begin{tabular}{lllllllllllllll}
\hline\hline
$\nu_{\rm mod}$&$(\ell,n_{\rm p}, n_{\rm g})$ &$\beta_{\ell, n}$ &$\nu_{\rm mod}$&$(\ell,n_{\rm p}, n_{\rm g})$ &$\beta_{\ell, n}$ &$\nu_{\rm mod}$&$(\ell,n_{\rm p}, n_{\rm g})$ &$\beta_{\ell, n}$ &$\nu_{\rm mod}$&$(\ell,n_{\rm p}, n_{\rm g})$ &$\beta_{\ell, n}$\\
\hline
196.2583  &(0,0,   0)  &       &  9.0610  &(1,0, -32)  &0.503  &  3.7467  &(2,0,-137)  &0.834  &  7.6076   &(2,0, -67)  &0.834\\
253.3330  &(0,1,   0)  &       &  9.3495  &(1,0, -31)  &0.504  &  3.7740  &(2,0,-136)  &0.834  &  7.7188   &(2,0, -66)  &0.834\\
311.0961  &(0,2,   0)  &       &  9.6595  &(1,0, -30)  &0.504  &  3.8019  &(2,0,-135)  &0.834  &  7.8321   &(2,0, -65)  &0.834\\
368.6966  &(0,3,   0)  &       &  9.9826  &(1,0, -29)  &0.504  &  3.8306  &(2,0,-134)  &0.834  &  7.9490   &(2,0, -64)  &0.834\\
          &            &       & 10.3010  &(1,0, -28)  &0.504  &  3.8598  &(2,0,-133)  &0.834  &  8.0712   &(2,0, -63)  &0.834\\
  3.0217  &(1,0, -98)  &0.500  & 10.6122  &(1,0, -27)  &0.505  &  3.8895  &(2,0,-132)  &0.834  &  8.1994   &(2,0, -62)  &0.834\\
  3.0525  &(1,0, -97)  &0.500  & 10.9678  &(1,0, -26)  &0.505  &  3.9193  &(2,0,-131)  &0.834  &  8.3331   &(2,0, -61)  &0.834\\
  3.0840  &(1,0, -96)  &0.500  & 11.3892  &(1,0, -25)  &0.505  &  3.9492  &(2,0,-130)  &0.834  &  8.4716   &(2,0, -60)  &0.834\\
  3.1164  &(1,0, -95)  &0.500  & 11.8657  &(1,0, -24)  &0.505  &  3.9793  &(2,0,-129)  &0.834  &  8.6147   &(2,0, -59)  &0.834\\
  3.1498  &(1,0, -94)  &0.500  & 12.3912  &(1,0, -23)  &0.506  &  4.0100  &(2,0,-128)  &0.834  &  8.7615   &(2,0, -58)  &0.834\\
  3.1840  &(1,0, -93)  &0.500  & 12.9617  &(1,0, -22)  &0.506  &  4.0414  &(2,0,-127)  &0.834  &  8.9119   &(2,0, -57)  &0.835\\
  3.2188  &(1,0, -92)  &0.500  & 13.5628  &(1,0, -21)  &0.506  &  4.0736  &(2,0,-126)  &0.834  &  9.0671   &(2,0, -56)  &0.835\\
  3.2542  &(1,0, -91)  &0.500  & 14.1473  &(1,0, -20)  &0.506  &  4.1065  &(2,0,-125)  &0.834  &  9.2289   &(2,0, -55)  &0.835\\
  3.2899  &(1,0, -90)  &0.501  & 14.7116  &(1,0, -19)  &0.507  &  4.1400  &(2,0,-124)  &0.834  &  9.3980   &(2,0, -54)  &0.835\\
  3.3263  &(1,0, -89)  &0.501  & 15.4063  &(1,0, -18)  &0.507  &  4.1738  &(2,0,-123)  &0.834  &  9.5742   &(2,0, -53)  &0.835\\
  3.3635  &(1,0, -88)  &0.501  & 16.2674  &(1,0, -17)  &0.508  &  4.2078  &(2,0,-122)  &0.834  &  9.7576   &(2,0, -52)  &0.835\\
  3.4017  &(1,0, -87)  &0.501  & 17.2756  &(1,0, -16)  &0.509  &  4.2420  &(2,0,-121)  &0.834  &  9.9474   &(2,0, -51)  &0.835\\
  3.4412  &(1,0, -86)  &0.501  & 18.4364  &(1,0, -15)  &0.510  &  4.2766  &(2,0,-120)  &0.834  & 10.1442   &(2,0, -50)  &0.835\\
  3.4819  &(1,0, -85)  &0.501  & 19.7688  &(1,0, -14)  &0.511  &  4.3119  &(2,0,-119)  &0.834  & 10.3493   &(2,0, -49)  &0.835\\
  3.5236  &(1,0, -84)  &0.501  & 21.2921  &(1,0, -13)  &0.512  &  4.3483  &(2,0,-118)  &0.834  & 10.5637   &(2,0, -48)  &0.835\\
  3.5659  &(1,0, -83)  &0.501  & 22.9950  &(1,0, -12)  &0.512  &  4.3856  &(2,0,-117)  &0.834  & 10.7871   &(2,0, -47)  &0.835\\
  3.6084  &(1,0, -82)  &0.501  & 24.7207  &(1,0, -11)  &0.511  &  4.4237  &(2,0,-116)  &0.834  & 11.0194   &(2,0, -46)  &0.835\\
  3.6514  &(1,0, -81)  &0.501  & 26.4240  &(1,0, -10)  &0.512  &  4.4622  &(2,0,-115)  &0.834  & 11.2604   &(2,0, -45)  &0.835\\
  3.6955  &(1,0, -80)  &0.501  & 28.8158  &(1,0,  -9)  &0.515  &  4.5013  &(2,0,-114)  &0.834  & 11.5103   &(2,0, -44)  &0.835\\
  3.7412  &(1,0, -79)  &0.501  & 32.1554  &(1,0,  -8)  &0.517  &  4.5407  &(2,0,-113)  &0.834  & 11.7706   &(2,0, -43)  &0.835\\
  3.7887  &(1,0, -78)  &0.501  & 36.6016  &(1,0,  -7)  &0.518  &  4.5806  &(2,0,-112)  &0.834  & 12.0435   &(2,0, -42)  &0.835\\
  3.8377  &(1,0, -77)  &0.501  & 42.5975  &(1,0,  -6)  &0.519  &  4.6214  &(2,0,-111)  &0.834  & 12.3321   &(2,0, -41)  &0.835\\
  3.8879  &(1,0, -76)  &0.501  & 50.9201  &(1,0,  -5)  &0.519  &  4.6632  &(2,0,-110)  &0.834  & 12.6393   &(2,0, -40)  &0.836\\
  3.9389  &(1,0, -75)  &0.501  & 63.0243  &(1,0,  -4)  &0.515  &  4.7059  &(2,0,-109)  &0.834  & 12.9664   &(2,0, -39)  &0.836\\
  3.9904  &(1,0, -74)  &0.501  & 82.1270  &(1,0,  -3)  &0.503  &  4.7496  &(2,0,-108)  &0.834  & 13.3124   &(2,0, -38)  &0.836\\
  4.0425  &(1,0, -73)  &0.501  &116.5050  &(1,0,  -2)  &0.485  &  4.7941  &(2,0,-107)  &0.834  & 13.6739   &(2,0, -37)  &0.836\\
  4.0959  &(1,0, -72)  &0.501  &189.9109  &(1,0,  -1)  &0.565  &  4.8391  &(2,0,-106)  &0.834  & 14.0445   &(2,0, -36)  &0.836\\
  4.1514  &(1,0, -71)  &0.501  &203.3093  &(1,1,  -1)  &0.898  &  4.8757  &(2,0,-105)  &0.835  & 14.4167   &(2,0, -35)  &0.836\\
  4.2092  &(1,0, -70)  &0.501  &262.1546  &(1,1,   0)  &0.994  &  4.8847  &(2,0,-104)  &0.834  & 14.7956   &(2,0, -34)  &0.836\\
  4.2692  &(1,0, -69)  &0.501  &325.5027  &(1,2,   0)  &0.996  &  4.9313  &(2,0,-103)  &0.834  & 15.2053   &(2,0, -33)  &0.836\\
  4.3311  &(1,0, -68)  &0.501  &389.7996  &(1,3,   0)  &0.994  &  4.9788  &(2,0,-102)  &0.834  & 15.6602   &(2,0, -32)  &0.837\\
  4.3944  &(1,0, -67)  &0.501  &          &            &       &  5.0273  &(2,0,-101)  &0.834  & 16.1576   &(2,0, -31)  &0.837\\
  4.4588  &(1,0, -66)  &0.501  &  3.0020  &(2,0,-171)  &0.833  &  5.0771  &(2,0,-100)  &0.834  & 16.6881   &(2,0, -30)  &0.837\\
  4.5243  &(1,0, -65)  &0.501  &  3.0195  &(2,0,-170)  &0.833  &  5.1281  &(2,0, -99)  &0.834  & 17.2333   &(2,0, -29)  &0.837\\
  4.5917  &(1,0, -64)  &0.501  &  3.0373  &(2,0,-169)  &0.833  &  5.1799  &(2,0, -98)  &0.834  & 17.7636   &(2,0, -28)  &0.837\\
  4.6623  &(1,0, -63)  &0.501  &  3.0555  &(2,0,-168)  &0.833  &  5.2325  &(2,0, -97)  &0.834  & 18.3035   &(2,0, -27)  &0.838\\
  4.7363  &(1,0, -62)  &0.501  &  3.0740  &(2,0,-167)  &0.833  &  5.2859  &(2,0, -96)  &0.834  & 18.9366   &(2,0, -26)  &0.838\\
  4.8135  &(1,0, -61)  &0.501  &  3.0927  &(2,0,-166)  &0.833  &  5.3404  &(2,0, -95)  &0.834  & 19.6736   &(2,0, -25)  &0.838\\
  4.8935  &(1,0, -60)  &0.501  &  3.1116  &(2,0,-165)  &0.833  &  5.3966  &(2,0, -94)  &0.834  & 20.4980   &(2,0, -24)  &0.838\\
  4.9762  &(1,0, -59)  &0.501  &  3.1305  &(2,0,-164)  &0.833  &  5.4543  &(2,0, -93)  &0.834  & 21.4025   &(2,0, -23)  &0.839\\
  5.0611  &(1,0, -58)  &0.501  &  3.1495  &(2,0,-163)  &0.833  &  5.5135  &(2,0, -92)  &0.834  & 22.3802   &(2,0, -22)  &0.839\\
  5.1481  &(1,0, -57)  &0.501  &  3.1686  &(2,0,-162)  &0.833  &  5.5738  &(2,0, -91)  &0.834  & 23.4012   &(2,0, -21)  &0.839\\
  5.2378  &(1,0, -56)  &0.501  &  3.1881  &(2,0,-161)  &0.833  &  5.6349  &(2,0, -90)  &0.834  & 24.3823   &(2,0, -20)  &0.839\\
  5.3313  &(1,0, -55)  &0.501  &  3.2080  &(2,0,-160)  &0.833  &  5.6968  &(2,0, -89)  &0.834  & 25.3580   &(2,0, -19)  &0.840\\
  5.4291  &(1,0, -54)  &0.501  &  3.2283  &(2,0,-159)  &0.834  &  5.7597  &(2,0, -89)  &0.834  & 26.5740   &(2,0, -18)  &0.841\\
  5.5310  &(1,0, -53)  &0.501  &  3.2490  &(2,0,-158)  &0.834  &  5.8240  &(2,0, -88)  &0.834  & 28.0605   &(2,0, -17)  &0.842\\
  5.6370  &(1,0, -52)  &0.501  &  3.2699  &(2,0,-157)  &0.834  &  5.8903  &(2,0, -87)  &0.834  & 29.7907   &(2,0, -16)  &0.842\\
  5.7468  &(1,0, -51)  &0.501  &  3.2908  &(2,0,-156)  &0.834  &  5.9587  &(2,0, -86)  &0.834  & 31.7772   &(2,0, -15)  &0.843\\
  5.8606  &(1,0, -50)  &0.501  &  3.3118  &(2,0,-155)  &0.834  &  6.0291  &(2,0, -85)  &0.834  & 34.0521   &(2,0, -14)  &0.844\\
  5.9793  &(1,0, -49)  &0.502  &  3.3329  &(2,0,-154)  &0.834  &  6.1012  &(2,0, -84)  &0.834  & 36.6456   &(2,0, -13)  &0.845\\
  6.1033  &(1,0, -48)  &0.502  &  3.3545  &(2,0,-153)  &0.834  &  6.1742  &(2,0, -83)  &0.834  & 39.5333   &(2,0, -12)  &0.845\\
  6.2326  &(1,0, -47)  &0.502  &  3.3765  &(2,0,-152)  &0.834  &  6.2478  &(2,0, -82)  &0.834  & 42.4434   &(2,0, -11)  &0.844\\
  6.3670  &(1,0, -46)  &0.502  &  3.3991  &(2,0,-151)  &0.834  &  6.3222  &(2,0, -81)  &0.834  & 45.3309   &(2,0, -10)  &0.845\\
  6.5066  &(1,0, -45)  &0.502  &  3.4221  &(2,0,-150)  &0.834  &  6.3985  &(2,0, -80)  &0.834  & 49.3875   &(2,0,  -9)  &0.848\\
  6.6513  &(1,0, -44)  &0.502  &  3.4453  &(2,0,-149)  &0.834  &  6.4776  &(2,0, -79)  &0.834  & 55.0127   &(2,0,  -8)  &0.850\\
  6.8019  &(1,0, -43)  &0.502  &  3.4687  &(2,0,-148)  &0.834  &  6.5598  &(2,0, -78)  &0.834  & 62.4566   &(2,0,  -7)  &0.851\\
  6.9599  &(1,0, -42)  &0.502  &  3.4921  &(2,0,-147)  &0.834  &  6.6445  &(2,0, -77)  &0.834  & 72.4121   &(2,0,  -6)  &0.852\\
  7.1269  &(1,0, -41)  &0.502  &  3.5156  &(2,0,-146)  &0.834  &  6.7314  &(2,0, -76)  &0.834  & 86.0599   &(2,0,  -5)  &0.851\\
  7.3048  &(1,0, -40)  &0.502  &  3.5395  &(2,0,-145)  &0.834  &  6.8195  &(2,0, -75)  &0.834  &105.4720   &(2,0,  -4)  &0.844\\
  7.4943  &(1,0, -39)  &0.502  &  3.5640  &(2,0,-144)  &0.834  &  6.9084  &(2,0, -74)  &0.834  &134.5426   &(2,0,  -3)  &0.821\\
  7.6950  &(1,0, -38)  &0.502  &  3.5892  &(2,0,-143)  &0.834  &  6.9984  &(2,0, -73)  &0.834  &176.8873   &(2,0,  -2)  &0.804\\
  7.9052  &(1,0, -37)  &0.503  &  3.6148  &(2,0,-142)  &0.834  &  7.0908  &(2,0, -72)  &0.834  &208.5777   &(2,0,  -1)  &0.982\\
  8.1217  &(1,0, -36)  &0.503  &  3.6409  &(2,0,-141)  &0.834  &  7.1870  &(2,0, -71)  &0.834  &250.1862   &(2,1,  -1)  &0.894\\
  8.3402  &(1,0, -35)  &0.503  &  3.6672  &(2,0,-140)  &0.834  &  7.2871  &(2,0, -70)  &0.834  &289.4570   &(2,2,  -1)  &0.868\\
  8.5616  &(1,0, -34)  &0.503  &  3.6936  &(2,0,-139)  &0.834  &  7.3910  &(2,0, -69)  &0.834  &313.7516   &(2,2,   0)  &0.901\\
  8.7984  &(1,0, -33)  &0.503  &  3.7200  &(2,0,-138)  &0.834  &  7.4981  &(2,0, -68)  &0.834  &357.6327   &(2,3,   0)  &0.887\\
\hline
\end{tabular}
\end{table*}
\begin{table*}
\centering
\caption{\label{t5}Comparison between model frequencies of the best-fitting model and the $\delta$ Scuti frequencies. $\nu_{\rm obs}$ is the observed frequency and $\nu_{\rm mod}$ is its corresponding model frequency. ($\ell, n_p, n_g, m$) are the spherical harmonic degree, the radial orders in p-mode propagation zone, the radial orders in g-mode propagation zone, and the azimuthal number of the model frequency, respectively. $|\nu_{\rm obs} - \nu_{\rm mod}|$ represents the frequency difference between the observed frequency and its model counterpart.}
\begin{tabular}{ccclc}
\hline\hline
ID         &$\nu_{\rm obs}$   &$\nu_{\rm mod}$ &($\ell$, $n_{\rm p}$, $n_{\rm g}$, $m$) & $|\nu_{\rm obs}-\nu_{\rm mod}|$\\
           &($\mu$Hz)         &($\mu$Hz)       &                                        &($\mu$Hz)\\
\hline
$F$        &196.531              &196.258           &(0, 0, 0, 0)        &0.273\\
$p_{1}$    &188.113              &188.316           &(1, 0,-1,-1)        &0.203\\
$p_{2}$    &253.138              &253.333           &(0, 1, 0, 0)        &0.195\\
$p_{3}$    &203.149              &203.033           &(2, 0,-1,-2)        &0.116\\
$p_{4}$    &250.061              &250.186           &(2, 1,-1, 0)        &0.125\\
$p_{5}$    &203.326              &203.309           &(1, 1,-1, 0)        &0.017\\
$p_{6}$    &308.669              &308.665           &(2, 2, 0,-2)        &0.004\\
$p_{7}$    &252.825              &252.710           &(2, 1,-1,+1)        &0.115\\
\hline
\end{tabular}
\end{table*}
\begin{table*}
\centering
\tiny
\caption{\label{t6}Comparison between model frequencies of the best-fitting model and the $\gamma$ Dor frequencies. $\nu_{\rm obs}$ and $P_{\rm obs}$ represent the frequency and period of the observed modes, respectively.
$P_{\rm mod}$ is the period of its model counterpart. ($\ell, n_p, n_g, m$) are the spherical harmonic degree, the radial orders in p-mode propagation zone, the radial orders in g-mode propagation zone, and the azimuthal number of the model frequency, respectively. $|P_{\rm obs} - P_{\rm mod}|$ represents the period difference between the observation and its model counterpart.}
\begin{tabular}{cccccccccccccc}
\hline\hline
ID  &$\nu_{\rm obs}$   &$P_{\rm obs}$ &$P_{\rm mod}$ &($\ell$, $n_{\rm p}$, $n_{\rm g}$, $m$) & $|P_{\rm obs}-P_{\rm mod}|$ &ID  &$\nu_{\rm obs}$   &$P_{\rm obs}$ &$P_{\rm mod}$ &($\ell$, $n_{\rm p}$, $n_{\rm g}$, $m$) & $|P_{\rm obs}-P_{\rm mod}|$\\
 &(d$^{-1}$)  &(days)  &(days)  &   &(days) &  &(d$^{-1}$) &(days)  &(days)   &   &(days)\\
\hline
\boldmath{$f_{1}$}  &1.5954  &0.6268  &0.6278   &(1,0,-15, 0)    &0.0010  &\boldmath{$f_{33}$}  &0.8183  &1.2220 &1.2237 &(2,0,-108, 2)   &0.0017\\
\boldmath{$f_{2}$}  &1.4342  &0.6973  &0.6985   &(2,0,-26,-1)    &0.0012  &$f_{34}$             &0.5610  &1.7825 &1.7822 &(2,0,-124, 1)   &0.0003\\
\boldmath{$f_{3}$}  &1.3660  &0.7321  &0.7307   &(1,0,-15,-1)    &0.0014  &$f_{35}$             &3.6095  &0.2770 &0.2761 &(2,0,-12,  1)   &0.0009\\
\boldmath{$f_{4}$}  &1.6903  &0.5916  &0.5906   &(2,0,-29, 1)    &0.0010  &$f_{36}$             &0.5274  &1.8961 &1.8964 &(1,0,-48,  0)   &0.0003\\
\boldmath{$f_{5}$}  &1.7974  &0.5564  &0.5542   &(2,0,-31, 2)    &0.0022  &$f_{37}$             &0.3865  &2.5873 &2.5921 &(2,0,-75, -1)   &0.0048\\
\boldmath{$f_{6}$}  &1.2488  &0.8008  &0.7998   &(2,0,-52, 2)    &0.0010  &\boldmath{$f_{38}$}  &2.6154  &0.3824 &0.3824 &(1,0,-9,   1)   &0.0000\\
\boldmath{$f_{7}$}  &1.1984  &0.8344  &0.8346   &(2,0,-44, 1)    &0.0002  &\boldmath{$f_{39}$}  &1.3037  &0.7670 &0.7683 &(2,0,-49,  2)   &0.0013\\
\boldmath{$f_{8}$}  &1.9165  &0.5218  &0.5252   &(2,0,-25, 1)    &0.0034  &$f_{40}$             &0.6158  &1.6239 &1.6240 &(1,0,-41,  0)   &0.0001\\
$f_{9}$             &0.4515  &2.2148  &2.2119   &(2,0,-97, 0)    &0.0029  &$f_{41}$             &0.7402  &1.3510 &1.3508 &(2,0,-133, 2)   &0.0002\\
\boldmath{$f_{10}$} &0.8973  &1.1145  &1.1123   &(2,0,-89, 2)    &0.0022  &$f_{42}$             &0.4068  &2.4582 &2.4595 &(2,0,-109, 0)   &0.0013\\
\boldmath{$f_{11}$} &2.2015  &0.4542  &0.4564   &(2,0,-19, 0)    &0.0022  &$f_{43}$             &0.5713  &1.7504 &1.7546 &(2,0,-121, 1)   &0.0042\\
$f_{12}$            &0.7279  &1.3738  &1.3732   &(2,0,-138,2)    &0.0006  &$f_{44}$             &0.3716  &2.6911 &2.6978 &(2,0,-77, -1)   &0.0067\\
$f_{13}$            &0.7501  &1.3332  &1.3340   &(2,0,-81, 1)    &0.0008  &$f_{45}$             &0.3969  &2.5195 &2.5186 &(1,0,-92,  1)   &0.0009\\
$f_{14}$            &0.4298  &2.3267  &2.3259   &(1,0,-59, 0)    &0.0008  &\boldmath{$f_{46}$}  &1.5079  &0.6632 &0.6667 &(2,0,-40,  2)   &0.0035\\
$f_{15}$            &0.4600  &2.1739  &2.1710   &(1,0,-55, 0)    &0.0029  &$f_{47}$             &2.4495  &0.4082 &0.4113 &(2,0,-21,  2)   &0.0031\\
\boldmath{$f_{16}$} &0.8429  &1.1864  &1.1862   &(2,0,-52, 0)    &0.0002  &$f_{48}$             &1.8449  &0.5420 &0.5408 &(2,0,-23,  0)   &0.0012\\
$f_{17}$            &0.7658  &1.3058  &1.3036   &(1,0,-28,-1)    &0.0022  &$f_{49}$             &0.3236  &3.0902 &3.0891 &(2,0,-137, 0)   &0.0011\\
$f_{18}$            &0.9521  &1.0503  &1.0503   &(2,0,-46, 0)    &0.0000  &\boldmath{$f_{50}$}  &2.4030  &0.4161 &0.4153 &(1,0,-10,  1)   &0.0008\\
$f_{19}$            &0.3067  &3.2605  &3.2699   &(2,0,-145,0)    &0.0094  &$f_{51}$             &0.4693  &2.1308 &2.1316 &(2,0,-50, -2)   &0.0008\\
\boldmath{$f_{20}$} &2.0490  &0.4880  &0.4890   &(2,0,-26, 2)    &0.0010  &$f_{52}$             &0.2995  &3.3389 &3.3367 &(2,0,-148, 0)   &0.0022\\
$f_{21}$            &0.4440  &2.2523  &2.2501   &(2,0,-68,-1)    &0.0022  &$f_{53}$             &1.6708  &0.5985 &0.5931 &(2,0,-34,  2)   &0.0054\\
$f_{22}$            &0.3118  &3.2072  &3.2075   &(1,0,-82, 0)    &0.0003  &$f_{54}$             &0.5381  &1.8584 &1.8585 &(1,0,-61,  1)   &0.0001\\
\boldmath{$f_{23}$} &1.1048  &0.9051  &0.9056   &(2,0,-63, 2)    &0.0005  &\boldmath{$f_{55}$}  &1.0641  &0.9398 &0.9397 &(2,0,-67,  2)   &0.0001\\
$f_{24}$            &0.6325  &1.5810  &1.5792   &(2,0,-42,-2)    &0.0018  &$f_{56}$             &3.6421  &0.2746 &0.2727 &(2,0,-11,  0)   &0.0019\\
$f_{25}$            &0.4757  &2.1022  &2.1005   &(1,0,-72, 1)    &0.0017  &$f_{57}$             &3.0996  &0.3226 &0.3177 &(2,0,-14,  1)   &0.0049\\
\boldmath{$f_{26}$} &0.9242  &1.0820  &1.0830   &(2,0,-61, 1)    &0.0010  &$f_{58}$             &0.3385  &2.9542 &2.9548 &(1,0,-55, -1)   &0.0006\\
\boldmath{$f_{27}$} &0.6987  &1.4312  &1.4315   &(2,0,-152,2)    &0.0003  &$f_{59}$             &0.6639  &1.5063 &1.5041 &(1,0,-38,  0)   &0.0022\\
$f_{28}$            &0.8367  &1.1952  &1.1949   &(2,0,-42,-1)    &0.0003  &$f_{60}$             &0.8310  &1.2034 &1.2006 &(2,0,-103, 2)   &0.0028\\
\boldmath{$f_{29}$} &0.6826  &1.4650  &1.4641   &(1,0,-37, 0)    &0.0009  &$f_{61}$             &2.4595  &0.4066 &0.4017 &(1,0,-9,   0)   &0.0049\\
$f_{30}$            &0.8004  &1.2494  &1.2495   &(2,0,-74, 1)    &0.0001  &$f_{62}$             &0.6064  &1.6491 &1.6493 &(2,0,-110, 1)   &0.0002\\
$f_{31}$            &0.5989  &1.6697  &1.6689   &(2,0,-112,1)    &0.0008  &$f_{63}$             &0.4968  &2.0129 &2.0140 &(1,0,-51,  0)   &0.0011\\
$f_{32}$            &1.1283  &0.8863  &0.8875   &(2,0,-61, 2)    &0.0012  \\
\hline
\end{tabular}
\end{table*}
\end{document}